# Three metallic BN polymorphs: 1D multi-threaded conduction in 3D network

Mei Xiong[a], Zhibin Gao[b], Kun Luo[c], FeiFei Ling[c], YuFei Gao[c], Chong Chen[a], Dongli Yu[c], Zhisheng Zhao[c], Shizhong Wei[a*]

[a] National Joint Engineering Research Center for Abrasion Control and Molding of Metal Materials, Henan University of Science and Technology, Luoyang 471003, China.

[b] Department of Physics, National University of Singapore, Singapore 117551, Republic of Singapore

[c] Center for High Pressure Science, State Key Laboratory of Metastable Materials Science and Technology, Yanshan University, Qinhuangdao 066004, China.

In this paper, three novel metallic $sp^2$/$sp^3$-hybridized Boron Nitride (BN) polymorphs are proposed by first-principles calculations. One of them, named as $t$P-BN, is predicted based on the evolutionary particle swarm structural search. $t$P-BN is constructed by two interlocked rings forming a tube-like 3D network. The stability and band structure calculations show $t$P-BN is metastable and metallic at zero pressure. Calculations for the density of states and electron orbits confirm that the metallicity originates from the $sp^2$-hybridized B and N atoms, and forming 1D linear conductive channels in the 3D network. According to the relationship between the atomic structure and electronic properties, another two 3D metastable metallic $sp^2$/$sp^3$-hybridized BN structures are constructed manually. Electronic properties calculations show that both of these structures have 1D conductive channel along different axes. The polymorphs predicted in this study enrich the structures and provide a different picture of the conductive mechanism of BN compounds.

# 1 Introduction

Boron Nitride (BN) compounds have exceptional thermal and chemical stabilities, and they are widely employed as semiconductors. Experimentally, there are three BN compounds, the bulked $sp^3$-hybridized structure cubic BN (cBN)[1] and wurtzite BN (wBN),[2] layered $sp^2$-hybridized hexagonal BN (hBN),[3] respectively. According to the structures of BN allotropes, cBN and wBN can be used as superhard materials, whereas hBN can be used as a lubricant.[3] Recent years have witnessed numerous breakthroughs in BN systems, including novel functional materials synthesis,[4-15] improvement of physical performance of cBN,[16,17] etc. Apart from the three BN compounds which we mentioned above, numerous other BN nanomaterials have been synthesized. For example, zero-dimensional (0D) nanocages,[4,5] one-dimensional (1D) nanotubes,[6-9] two-dimensional (2D) nanosheets,[7,10] and three-dimensional (3D) nanofoam.[11] Furthermore, the Vickers hardness of cBN has been raised to ~100 GPa via synthesizing nanotwinned cBN.[18]

Carbon is an isoelectronic structure of BN structure, some of these atomic structures and physical properties features of the two polymorphs are similar to each other. For example, both of bulked cBN and diamond are fully $sp^3$-hybridized structures and can be used as superhard materials, the layered hBN and graphite is used as a lubricant. However, the electronic properties of these two systems are different. Carbon allotropes can be conductive (such as graphite and synthesized carbon nanotubes)[19] and insulating (such as diamond) depending on their framework. To date, all of the synthesized BN materials are semiconductors. Both of the 1D BN nanotubes and 2D BN nanosheets are wide band-gap semiconductors, independent of their chirality and edges.[20-24] According to these similarities between BN and carbon allotropes, metallic structures might exist in the BN system. Notably, studies showed that some of the bare BNNRs are metals, but this property is unstable, due to the highly active edges.[25-27] Comparing to 0D, 1D and 2D structures, 3D bulked structures have distinct advantages in terms of their stability of physical and chemical properties.

Recently, an increasing number of BN structures are predicted with the aid of structure searching. For example, the superhard *bct*-, *z*-, *p*-BN, the fully *sp*$^2$-hybridized 3D networks, and so on.[28-35] All of these structures are semiconductors, and no metals were predicted until 2013. Zhang et al.[36] and Dai et al.[37] reported metallic BN allotropes in their studies respectively, which indicate that conductive BN compounds may be synthesized via suitable precursors and methods. Subsequently, Xie et al. predicted a metallic N-rich B-N compounds.[38] Interestingly, metallic B-N structures possess different conductive mechanisms. For example, in Zhang's work, a conductive network is formed parallel to (001) crystal surface of T-$B_3N_3$,[36] whereas in Dai's research, both the two structures possessing 3D conductive networks.[37] These interesting results inspired the metallic structures predictions in the B-N system.

In this work, three novel metastable 3D BN structures are predicted. One of them, named *t*P-BN is predicted by the recently developed ab initio evolutionary algorithm CALYPSO, which is helpful for crystal structure prediction studies. The *sp*$^2$-hybridized atoms form an interlayer structure, owing to the 1D conduction in 3D network. Because of the *sp*$^2$/*sp*$^3$-hybridized structure, the mechanical properties of this metallic structure such as Vickers hardness and strength are comparable to some ceramics. The other two are constructed with *sp*$^2$-/*sp*$^3$-hybridized atoms, forming a 3D network. The elastic constants meet the criteria for mechanical stability, and phonon frequencies calculations demonstrate the dynamical stability. The calculated band structure and density of states suggest the 1D conduction of these novel structures.

## 2 Computational methods

In this work, the structure search process was performed by the particle swarm optimization methodology implemented by the well-developed CALYPSO (Crystal structure Analysis by Particle Swarm Optimization) code,[39, 40] which had been used in many previous studies.[40, 41] The calculation was performed assuming at zero pressure, and the simulation cell changed from 4 to 30 atoms. Subsequent structural relaxations were conducted using density functional theory (DFT) with the Perdew−Burke−Ernzerh (PBE) generalized gradient approximation (GGA) and PAW potential, as implemented in the VASP code.[42, 43] The used convergence plane-wave cutoff energy was set to 550 eV. The properties calculations of the selected structures were performed using the CASTEP code.[44] These included energetic properties, phonon spectrum, electronic properties (band structure, projected density of states, electron orbits), and mechanical properties (Vickers hardness, tensile strength). The local density approximation (LDA) exchange-correlation function of Ceperley and Alder parameterized by Perdw and Zunger (CA-PZ) was used to describe the electron-electron exchange interaction.[45, 46] The cutoff energy and Monkhorst–Pack $k$ meshes were carefully chosen to achieve the total energy convergence of 1–2 meV per BN unit. We set the cutoff energy of 770 eV for the norm-conserving (NC) potential. The Monkhorst–Pack $k$ meshes of $2\pi \times 0.04$ Å$^{-1}$ separations was chosen for the properties calculations. The dynamic stability of the structures was verified by calculating the phonon frequencies through a finite displacement method. Both the LDA functional and Heyd−Scuseria−Ernzerhof hybrid functional (HSE06)[47] were employed to calculate the electronic properties of investigated $t$P-BN. The unit cell was used to calculate the phonon dispersion spectra, electronic properties, the elastic constants, bulk and shear modulus with 0.003 was selected as the applied maximum strain amplitude during the process.

## 3 Results and Discussion

### 3.1 $t$P-BN structure

#### 3.1.1 Structure characterization

During the structures searching, all of the experimental BN allotropes were found, and the previously known BN allotropes were reproduced, such as $bct$- and $p$-BN T-B$_3$N$_3$, etc.[29, 35, 36] Under zero pressure, the equilibrium lattice parameters of the novel BN structure are a = b = 2.53Å, c = 9.87 Å, α = β = γ = 90°, with five B-N pairs in per unit cell, assembled in a tetragonal lattice (space group P-4m2, No. 115), therefore named as $t$P-BN. Fig. 1 shows the structural feature of $t$P-BN. There are six inequivalent atomic positions in $t$P-BN framework. Three B atoms, B1, B2 and B3 atoms occupy the 1b (0.5, 0.5, 0), 2e (0, 0, -0.822), and 2g (0, -0.5, -0.59) position, respectively.

Three N atoms, N1, N2 and N3 atoms occupy the 2g (0, 0.5, -0.09), 2e (0, 0, -0.32), and 1c (0.5, 0.5, 0.5) position, respectively. The B2 and N2 atoms are $sp^2$-hybridized, whereas the others are $sp^3$-hybridized. As shown in Fig. 1a, *t*P-BN can be viewed as 12-atoms rings (the red rectangle) interlocked together to form a 3D network in the unit cell.

### 3.1.2 Stabilities

**Thermodynamic stability**

To examine the thermodynamic stability, the enthalpies of the *t*P-BN structure are calculated as a function of pressure are calculated, and compared with several theoretically proposed BN structures, such as *bct*-, *z*-BN, T-B$_3$N$_3$, as depicted in Fig. 2. Accurate structural optimizations were performed for these phases from 0 GPa to 60 GPa. From the energetic point of view, *t*P-BN is found to be metastable as compared to the proposed semiconductor *bct*-BN, *p*-BN, and metallic M-BN, but energetically more stable than the previously proposed metallic T-B$_3$N$_3$ structure, at zero as well as high pressures.

**Dynamic stability**

To evaluate the dynamic stability of tP-BN, we investigate the phonon dispersion and frequency density of states (DOS) in the first Brillouin zone at zero pressure. The coordinates of the high symmetry points sequence in reciprocal space are Z (0, 0, 0.5) → A (0.5, 0.5, 0.5) → M (0.5, 0.5, 0.0) → G (0, 0, 0) → Z (0, 0, 0.5) → R (0, 0.5, 0.5) → X (0, 0.5, 0) → G (0, 0, 0). As plotted in Fig. 3a, we find that there are no imaginary frequencies in the whole Brillouin zone, which indicates that *t*P-BN is dynamically stable. The highest calculated phonon frequency of bond vibrational modes in *t*P-BN is 38.1 THz, a significant phonon gap departs the whole dispersion into two parts, the low frequency (0-33.3 THz) and the high frequency (37.9-38.1 THz) regions. According to the phonon DOS results shown in Fig. 3b, the high-frequency modes are entirely contributed by the B2 and N2 atoms, while the partial DOS of $sp^3$-hybridized atoms and total DOS overlap in the low-frequency region.

**Mechanical stability**

The mechanical stability is investigated by elastic constants calculations. The results

are: $C_{11}$ = 709.2, $C_{33}$ = 928.0, $C_{44}$ = 177.0, $C_{66}$ = 127.5, $C_{12}$ = 68.6 and $C_{13}$ = 138.0 GPa, respectively. For a stable tetragonal structure, all six independent elastic constants should meet the generalized Born stability criteria:[48] $C_{11} > 0$, $C_{33} > 0$, $C_{44} > 0$, $C_{66} > 0$, $C_{11} - C_{12} > 0$, $C_{11} + C_{33} - 2C_{13} > 0$, $2(C_{11} + C_{12}) + C_{33} + 4C_{13} > 0$. These elastic constants satisfy the criteria, which indicate that $t$P-BN is mechanically stable. The calculated bulk modulus is 332.9 GPa, indicating that $t$P-BN is more compressible than cBN (400 GPa).[49]

### 3.1.3 Mechanical properties

**Strength**

To study the mechanical properties of the $t$P-BN structure, the tensile strength and theoretical Vickers Hardness are calculated. According to the tetragonal crystal structure of $t$P-BN, we choose the [100], [001], and [110] directions to determine the ideal tensile strength. As shown in Fig.4, the strength is investigated for the different percentages (5%, 10%, 15%, 20%....60%) of tensile strains. The calculated tensile strengths along [100], [001] and [110] are 66.47 GPa with a strain of 0.25, 125.86 GPa with a strain of 0.2, and 147.5 GPa with a strain of 0.55, respectively. Apparently, these results manifest the anisotropy of the tensile strengths of $t$P-BN. The [110] tensile strength 147.5 GPa is comparable to the in-plane tensile strength of graphene (130 ± 10 GPa).[50] The weakest tensile strength is along the [100] direction, which indicates that the crystal will cleave along [100] direction as soon as the external stress exceeds 66.47 GPa.

**Vickers Hardness**

Generally, the theoretical Vickers Hardness of structures can be calculated with the microscopic model and empirical formula reported by Tian et al.[51, 52] and Chen et al.,[53] respectively. Considering the $sp^2$-hybridized atoms and tube-like features of the $t$P-BN framework, the theoretical Vickers Hardness is calculated using the empirical formula. The formula is depicted as: $H_v$ (GPa)=$2(G^3/B^2)^{0.585} - 3$. The calculated bulk modulus (B), shear modulus (G) and Young's modulus (E) of $t$P-BN structure are 332.9 GPa, 214.0 GPa and 528.7 GPa, respectively. Consequently, the calculated Vickers Hardness value is 24.5 GPa, which is comparable to $\alpha$-Al$_2$O$_3$(~20 GPa)[54] ceramic, implying that $t$P-BN is a hard material.

### 3.1.4 Electronic properties

The electronic properties of $t$P-BN are studied by calculating its band structure, and the results are shown in Fig. 5. The high symmetry K-point path of band calculation we chose in the Brillouin Zone is the same as the phonon calculation. Firstly, the band structure calculated with DFT calculations is plotted with red lines in Fig. 5a. The results show that there are three partially occupied bands cross the Fermi level in the Brillouin Zone, indicating the metallic feature of $t$P-BN. However, DFT calculations are well-known to underestimate the band gap.[55] In terms of deriving a more accurate band structure, the screened hybrid functional HSE06 has been demonstrated to be more credible than DFT calculations. Thus, we re-calculated the band structure with the HSE06 functional, and the results are depicted in Fig. 5a, marked with blue lines. One can see that there are two partially occupied band-crossing the Fermi level, which confirms the inherent metallicity of the $t$P-BN.

As revealed in previous studies, metallic BN structures possess a unique conductive mechanism depending on their atomic framework.[36-38] The origin of metallicity requires further exploration, thus we first calculated the density of states (DOS) and projected density of states (PDOS) of $t$P-BN and plotted the results in Fig. 5b. As shown in Fig. 5b, the contributions at the Fermi level are originated from the p orbitals of these atoms. Moreover, as shown in Fig. 5c-5e, most of the electron states at the Fermi level are contributed by the p orbitals of N2 and B2 atoms, and the contributions from B2 atoms are smaller than N2 atoms. To further study the conductive mechanism, we calculated the electron orbits of $t$P-BN, and plotted the results in Fig. 5f and 5g. The drawn electron orbits are the summation of the partially band-crossing the Fermi level. As shown in Fig. 5f the $sp^2$-hybridized B2 and N2 atoms formed a layered configuration along the $x$- and $y$-axis, respectively. In the left part of Fig. 5f, the orbits of N2 atoms are overlapped along $y$-axis and the orbits of B2 atoms are overlapped along $x$-axis. The geometries in the right parts are opposite to the left, namely the orbits of N2 atoms are overlapped along $x$-axis, and the orbits of B2 atoms are overlapped along $y$-axis. Moreover, as shown in Fig. 5g, the conductive channels are separated from each other. This indicates the *t*P-BN is linearly conductive, with independent conductive channels in its framework, i.e., a 3D BN network with peculiar 1D dual-threaded conductivity. This revealed conductive mechanism is different from the previously predicted metallic BN structures, such as T-$B_3N_3$ which has a 2D conductive network,[36] the structures predicted by Dai have a 3D network,[37] and the t-$B_3N_4$ has a 2D conductive network.[38]

**3.2 Manually construction of metallic BN structures**

**3.2.1 Structure Construction**

The proposed metallic B-N structures demonstrate that the well-semiconductor BN can achieve metallicity depending on the atomic configuration.[36-38] Due to the hybridization diversity of B and N atoms, some other metallic structures may exist in B-N system. In terms of structure configuration, there are several similarities exist among T-$B_3N_3$, (P-6M2)-BN, (IMM2)-BN and *t*-$B_3N_4$, which include $sp^2$-hybridized atoms, flat layers and a smaller interlayer distance (~2.6 Å) than hBN (~3.33 Å) to form a tube-like 3D network. Moreover, both of *t*P-BN and T-$B_3N_3$ have interlocked rings that form the tube-like 3D structure. Therefore, we have constructed several BN structures by adding $sp^2$-hybridized B-N pairs based on the *t*P-BN framework, forming a tube-like 3D network to search metallic BN polymorphs. As a result, two metastable BN structures have been obtained at zero pressure, named as $B_8N_8$-I and $B_8N_8$-II, respectively. Under zero pressure, the equilibrium lattice parameters of the $B_8N_8$-I structure are a = b = 2.55 Å, c = 15.86 Å, with four B and four N atoms in per unit cell, assembled in a tetragonal lattice (space group P42mc, No. 105), the atomic positions are listed in Table S1 (see in ESI†). We have plotted the frameworks of $B_8N_8$-I in Fig. 6a and 6b. As shown in Fig. 6a, similarly to *t*P-BN, $B_8N_8$-I is constructed by the interlocked blocks marked with the red frame. Fig. 6b shows the $sp^2$/$sp^3$ hybridized features of $B_8N_8$-I, and the $sp^2$ hybridized atoms (B2, B3, N2, N3), and $sp^3$-hybridized atoms (B1, B4, N1, N4). Thereafter, by switching the atomic positions of B and N atoms in the $B_8N_8$-I structure, we obtained another framework, named as $B_8N_8$-II, that shown in Fig. 6c and 6d.

**3.2.2 Stability**

We tested the thermodynamic stability by calculating the total energy at zero pressure. Results show that $B_8N_8$-I and $B_8N_8$-II structure are ~0.22 eV per BN unit lower than that of $t$P-BN. Therefore, at zero pressure, the constructed $B_8N_8$-I and $B_8N_8$-II are thermodynamically more stable than $t$P-BN. The dynamic stability of the $B_8N_8$-I and $B_8N_8$-II structure is investigated via the phonon dispersion calculations at zero pressure, and the results are shown in Fig. S1 (see in ESI†). The coordinates of the high symmetry points sequence in the first Brillouin are the same as those of $t$P-BN. As shown in Fig. S1a and S1b, there are no imaginary frequencies in the whole Brillouin zone, which means that the two structures are dynamically stable. The mechanical stability of the two structures is investigated by elastic constants calculations, and the results are listed in Table S2 (see in ESI†). These elastic constants satisfy the criteria for a stable tetragonal structure, which indicate that both $B_8N_8$-I and $B_8N_8$-II are mechanically stable.

### 3.2.3 Electronic properties

In order to study the electronic properties of the two structures $B_8N_8$-I and $B_8N_8$-II, the band structures and PDOS are calculated. As we mentioned above, the band structure is calculated by the HSE06 and LDA functional respectively, to obtain reliable results (see Fig. S2a and S2b in ESI†). In Fig. S2a and S2b, the blue lines stand for the results calculated by LDA functional, the red lines depict the results derived by HSE06 functional. As shown by previous studies, the band crossing Fermi level and the obvious density of state peak at the Fermi level indicate the metallic feature of crystal structures.[56, 57] Fig. S2a shows the band structure and DOS of $B_8N_8$-I. Four partially occupied bands crossing the Fermi level in the Brillouin Zone (blue line) are observed, indicating the metallic feature of $B_8N_8$-I. The DOS of $B_8N_8$-I can be observed the occupations at the Fermi level, which is consistent with the result we derived from the band structure. The similar results can be observed in Fig. S2b, which is the band structure and DOS of $B_8N_8$-II. There are six partially occupied bands crossing the Fermi level in the band structure (blue line), and obvious density of state peak at the Fermi level in the DOS, which are indicating the metallic feature.

For $B_8N_8$-I structure, the PDOS curves in Fig. S2c (see ESI†) show that occupations at the Fermi level are mainly contributed by the *sp*$^2$-hybridized B3, N2 and N3 atoms. For $B_8N_8$-II structure, the partially DOS curves in Fig. S2d (see ESI†) show that most of the occupations at the Fermi level are contributed by the *sp*$^2$-hybridized N3, B2 and B3 atoms.

The electron orbits originated from the bands across the Fermi level of these two structures are also calculated to investigate the conductive mechanism. As shown in Fig. 6, each of the two structures can be divided into two parts, the left part including the atoms in the red frame, and the right part including the remaining atoms. The calculated results for electron orbits of $B_8N_8$-I originated from the bands across the Fermi level are shown in Fig. 7a and 7b. In the left part of Fig. 7a, the electron orbits of the *sp*$^2$-hybridized B3 atoms are overlapped along *x*-axis, which are originated from the delocalized π electrons of B3-p orbitals. Similarly, the overlapped electron orbits of *sp*$^2$-hybridized N3 atoms can be seen between the N3 layers along *y*-axis. Fig. 7b shows the detail of blue frame in Fig. 7a, we can observe the electron orbits of *sp*$^2$-hybridized N2 atoms are connected together along the N2-B2-N2 chain. The electron orbits of the right part we calculated can be viewed as rotating the left part with 90°. That is the conductive channel formed by B3, N2 and N3 atoms is along *y*-axis, N2-B2-N2 chain and *x*-axis, respectively.

The underlying conductive mechanism is owing to the short interlayer distance between the *sp*$^2$-hybridized atoms along the axis. For example, the interlayer distance of B3 atoms is 2.56 Å, which is much smaller than the layer separation in hBN structure (~3.33 Å). Therefore, the π interactions between B3 atoms are enhanced and formed conductive channels. Moreover, the π electrons of B2 atoms are restricted by N2 and N3 atoms with the stronger electronegativity, inducing no π-like interactions between B2 atoms even if with a small interlayer separation. These results are consistent with the analyses of PDOS.

The conductive mechanism of $B_8N_8$-II structure is similar to that of $B_8N_8$-I structure. As shown in Fig. 7c, for the left part of $B_8N_8$-II, owing to the short interlayer distance (~2.56 Å), the π electrons of *sp*$^2$-hybridized N3, B2 and B3 atoms made the p orbitals of them overlapped along *x*-, *y*- and *y*-axis, respectively. Fig. 7d is the detail of the section in the blue frame of Fig. 7c. It shows the three conductive channels are independent, none of them are overlapped. In the right part, the electron orbits of N3, B2 and B3 atoms are overlapped along *y*-, *x*- and *x*-axis, respectively. Similarly, the three conductive channels also isolated to each other. However, there are no electron orbits overlapping between the $sp^2$-hybridized N2 atoms. This might be caused by the unique crystal configuration, which is the N1-B3 bonds restricted the electrons of N2 atoms.

All of the three novel BN polymorphs have 1D conductive channel in their 3D network, the differences between them lie in the counts of conductive channels, which are caused by the crystal structure. In the previous studies, the results show the electronic properties of structures depending on the atomic configurations.[56, 57] For example, the (3,3) BNT have four conductive channels, whereas (5,0) BNT have three conductive channels.[56] What's more, the metal-semiconductor transition can be induced by the distorted B atoms.[56] In this work, there are three, four and six partially occupied bands crossing the Fermi level in the Brillouin Zone of the *t*P-BN, $B_8N_8$-I and $B_8N_8$-II, respectively. Moreover, there are four conductive channels in the *t*P-BN, whereas six conductive channels in the $B_8N_8$-I and $B_8N_8$-II, respectively. With the $sp^2$-hybridized atoms adding, the counts of conductive channel increasing correspondingly, and the electron distributions are changed according to the configurations of these $sp^2$-hybridized atoms.

**Conclusion**

In sum, three metastable $sp^2/sp^3$-hybridized BN polymorphs are proposed as metallic frameworks and are demonstrated to be mechanically and dynamically stable at zero pressure. Structurally, these three BN structures possess a tetragonal unit cell with interlocked units to form a tube-like configuration. The band structure, DOS and electron orbits calculations show that $sp^2$-hybridized B and N atoms in these structures constitute independent 1D multi-threaded conductive channel along *x*- and *y*-axis, respectively. These results indicate that the $sp^2/sp^3$-hybridized BN structure with interlocked units can be taken into consideration during the BN metallic structure searching. The electronic properties of three BN structures demonstrate the diversity of the conductive mechanism of metallic BN and indicate their potential applications in electronic devices.

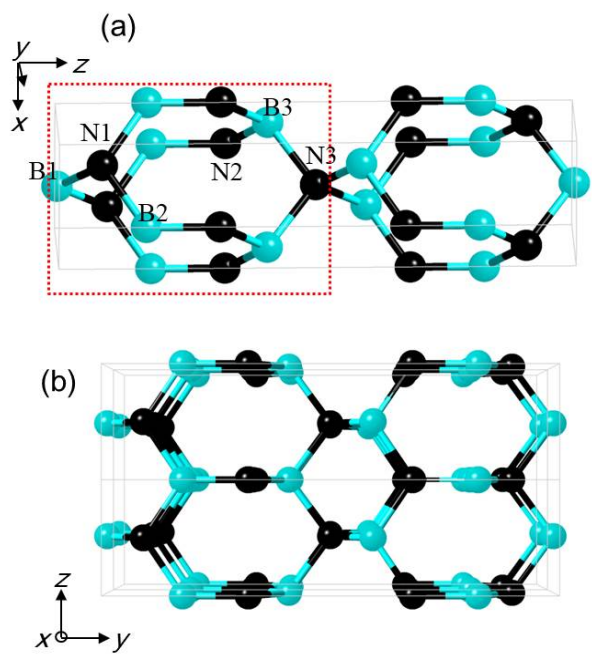

**Fig. 1** Atomic Configurations of *t*P-BN. (a) top view, (b) perspective view. The cyan and black balls stand for B and N atoms, respectively.

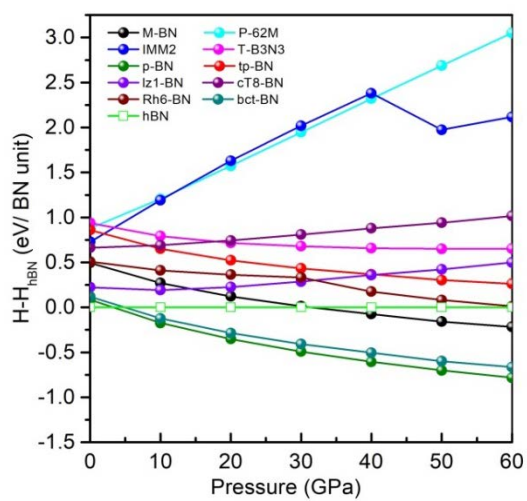

**Fig. 2**. Calculated enthalpies of BN structures relative to the hBN as a function of pressure.

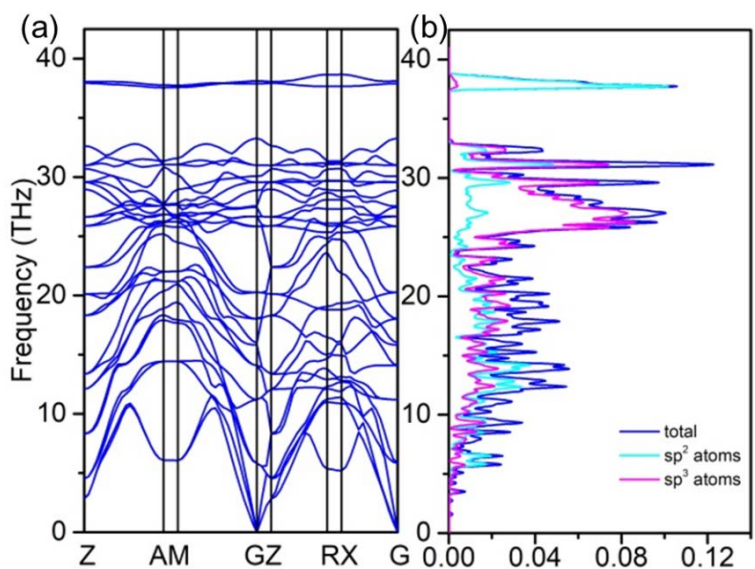

**Fig. 3** Phonon dispersion (a) and frequency DOS (THz$^{-1}$) of $t$P-BN.

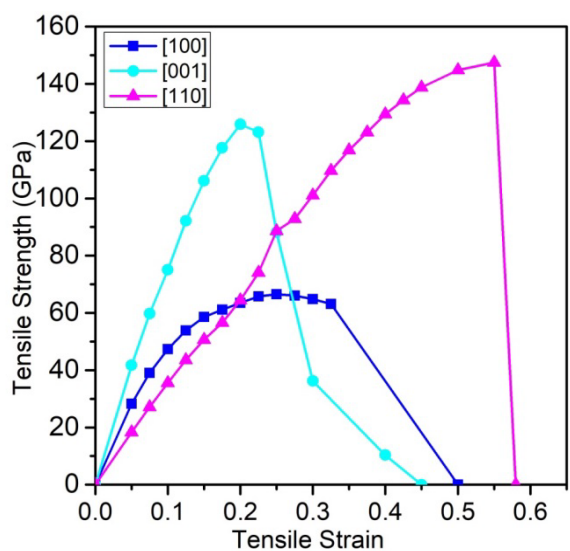

**Fig. 4** Calculated tensile strengths versus tensile strains of $t$P-BN along the [100], [001], and [110] directions, respectively.

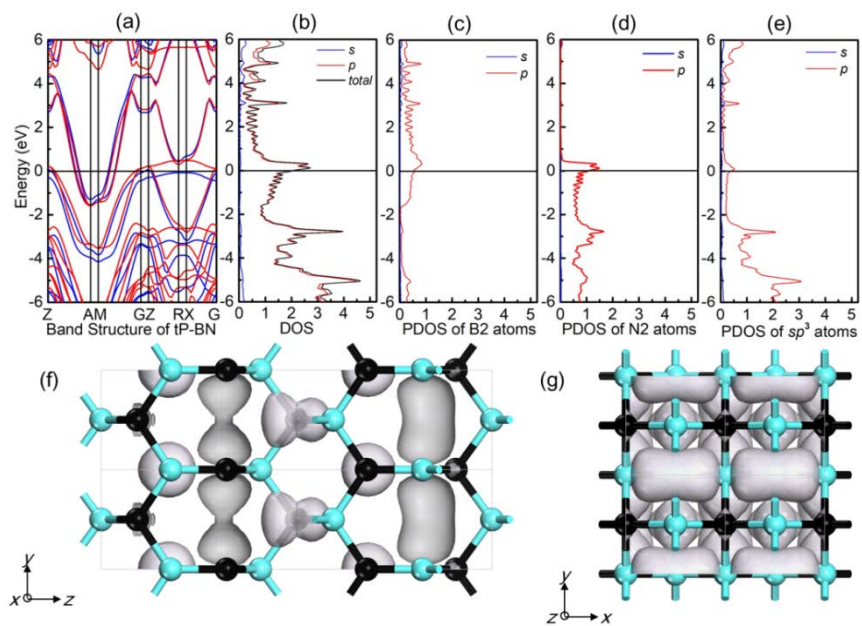

**Fig. 5** Electronic properties of $t$P-BN. (a) band structure, (b) the density of states, (c)-(e) projected density of states, (f)-(g) electron orbits of $t$P-BN across the Fermi level. The selected isosurface is 0.04. The cyan and black balls stand for B and N atoms, respectively.

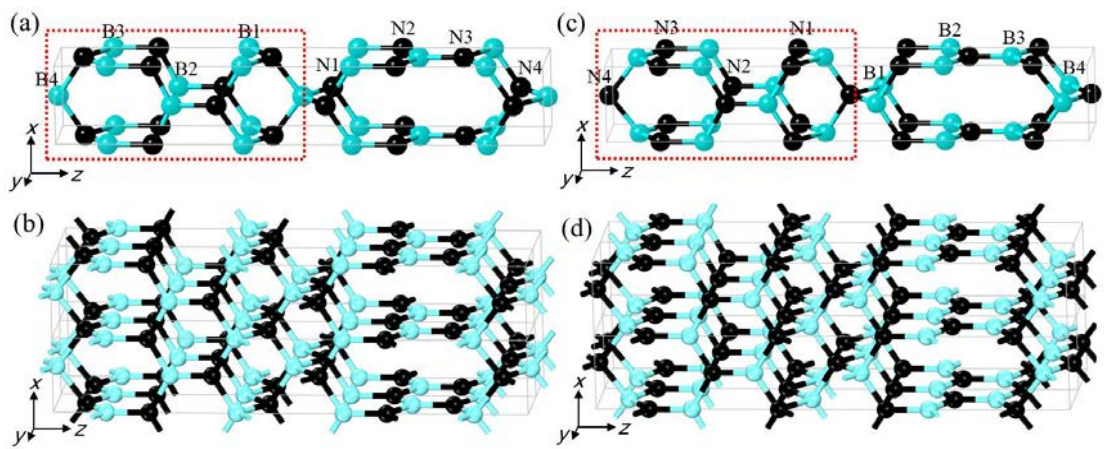

**Fig. 6** Structure configurations of constructed metallic BN structures. (a)-(b) $B_8N_8$-I, (c)-(d) $B_8N_8$-II. The cyan and black balls stand for B and N atoms, respectively.

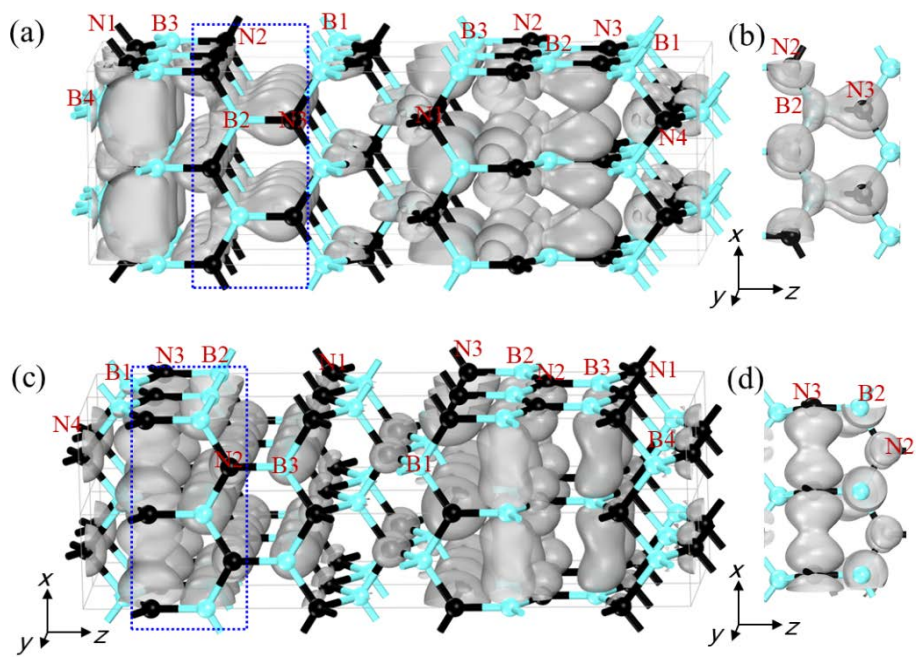

**Fig. 7** Electron orbits of (a)-(b) $B_8N_8$-I, and (c)-(d) $B_8N_8$-II across the Fermi level. The selected isosurface is 0.033. The cyan and black balls stand for B and N atoms, respectively.

# Electronic Supplementary Information

# Three metallic BN polymorphs: 1D multi-threaded conduction in 3D network

Mei Xiong[a], Zhibin Gao[b], Kun Luo[c], FeiFei Ling[c], YuFei Gao[c], Chong Chen[a], Dongli Yu[c], Zhisheng Zhao[c], Shizhong Wei[a*]

[a] National Joint Engineering Research Center for Abrasion Control and Molding of Metal Materials, Henan University of Science and Technology, Luoyang 471003, China.

[b] Department of Physics, National University of Singapore, Singapore 117551, Republic of Singapore

[c] Center for High Pressure Science, State Key Laboratory of Metastable Materials Science and Technology, Yanshan University, Qinhuangdao 066004, China.

**Fig. S1** Phonon dispersion of BN polymorphs.

**Fig. S2** Electronic band structures and density of states of BN polymorphs.

**Table S1** Space group (S.G.), lattice parameters (Å), and atomic Wyckoff positions of BN polymorphs at ambient pressure.

**Table S2** Elastic constants $C_{ij}$ (GPa), bulk modulus $B$ (GPa), and shear modulus $G$ (GPa) of BN polymorphs.

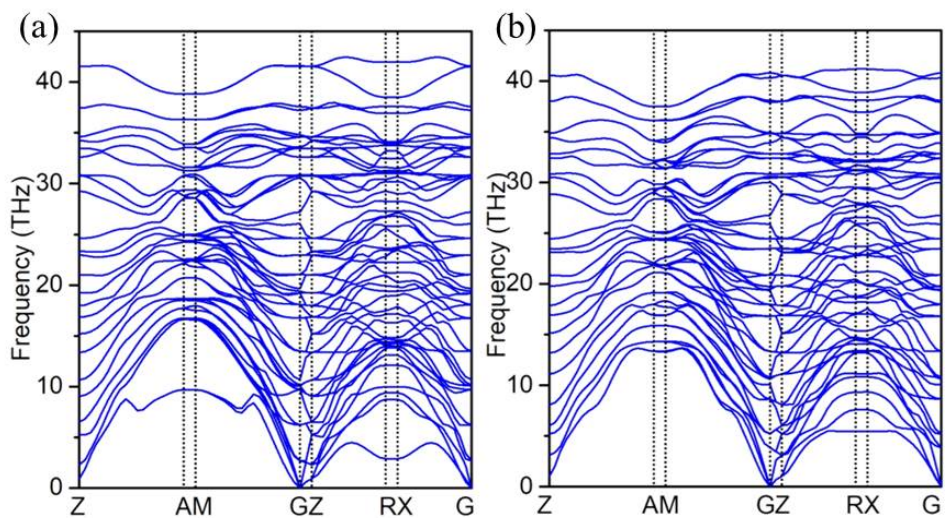

**Fig. S1** Phonon dispersion of (a) $B_8N_8$-I (b) $B_8N_8$-II at zero pressure.

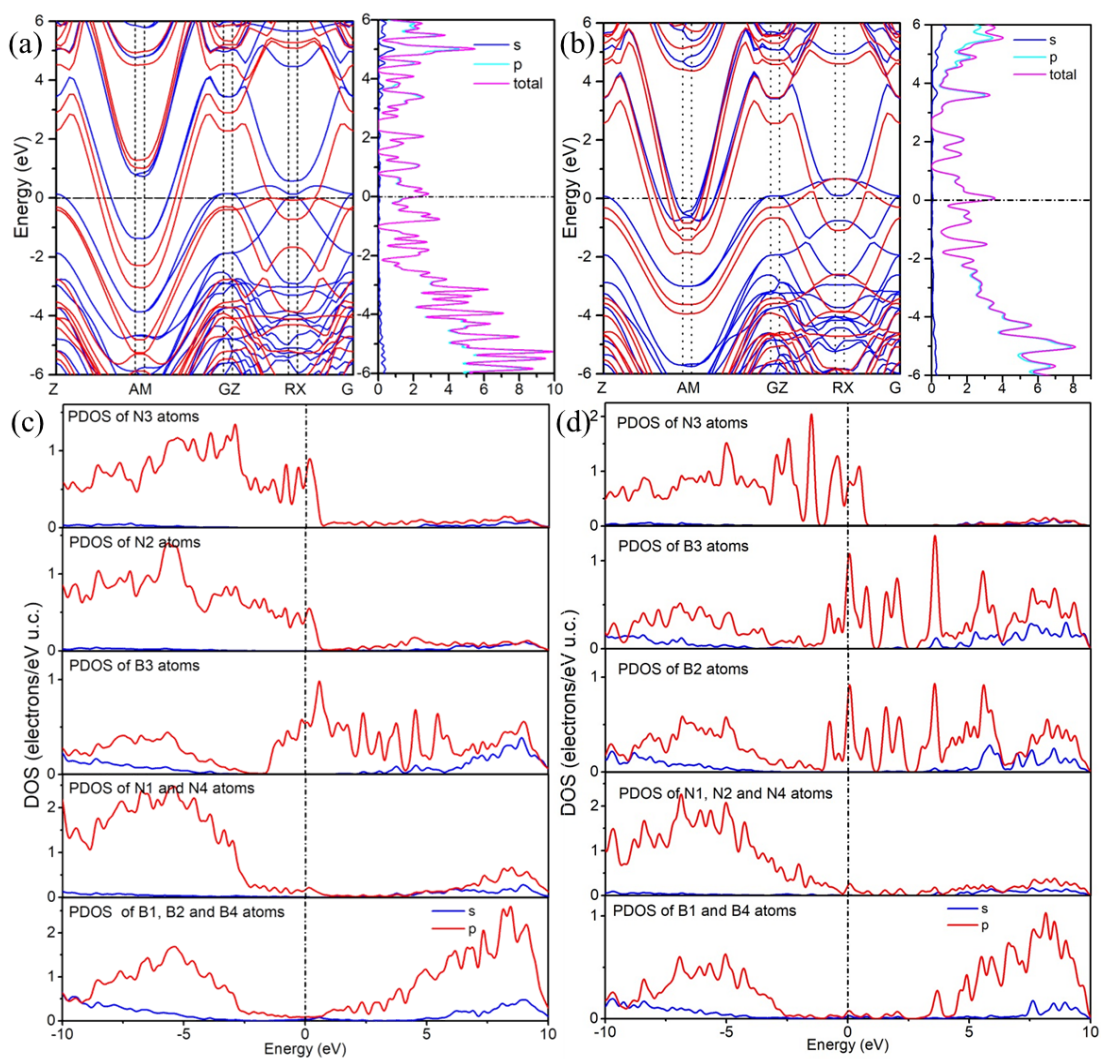

**Fig. S2** Electronic properties of onstructed metallic BN structures. (a) band structure of $B_8N_8$-I, (b) band structure of $B_8N_8$-II, (c) PDOS of $B_8N_8$-I, (d) PDOS of $B_8N_8$-II.

**Table S1.** Space group (S.G.), lattice parameters (Å), and atomic Wyckoff positions of BN polymorphs at ambient pressure.

| Structure | S.G. | lattice parameters | Atomic positions |
|---|---|---|---|
| $B_8N_8$-I | *P42mc* | a = b = 2.55 Å, c = 15.86 Å; $\alpha = \beta = \gamma = 90°$ | B1 2a (0,0,0.886), B2 2c (0,0.5,0.748), B3 2a (0,0,0.611); B4 2b (0.5,0.5,0.5); N1 2c (0.5,0,0.561); N2 2a (0,0,0.702); N3 2c (0,0.5,0.836); N4 2c (0.5,0,0.947) |
| $B_8N_8$-II | *P42mc* | a = b = 2.55 Å, c = 15.91 Å; $\alpha = \beta = \gamma = 90°$ | B1 2c (0.5,0,0.561); B2 2a (0,0,0.702); B3 2c (0,0.5,0.836); B4 2c (0.5,0,0.947); N1 2a (0,0,0.886), N2 2c (0,0.5,0.748), N3 2a (0,0,0.611); N4 2b (0.5,0.5,0.5); |

**Table S2.** Elastic constants $C_{ij}$ (GPa), bulk modulus $B$ (GPa), and shear modulus $G$ (GPa) of BN polymorphs.

| Structure | $C_{11}$ | $C_{33}$ | $C_{44}$ | $C_{66}$ | $C_{12}$ | $C_{13}$ | $B$ | $G$ |
|---|---|---|---|---|---|---|---|---|
| $B_8N_8$-I | 731.4 | 916.7 | 135.5 | 113.4 | 37.2 | 138.0 | 329.6 | 192.1 |
| $B_8N_8$-II | 695.7 | 898.8 | 122.5 | 109.7 | 52.2 | 149.6 | 327.4 | 178.9 |